\documentclass{cs19proc}

\editors{G.~A. Feiden}
\publisher{Zenodo}
\conference{The 19th Cambridge Workshop on Cool Stars, Stellar Systems, and the Sun}
\conferencedate{2016}

\title{Coronal activity cycles in action - X-rays from $\alpha$~Centauri A/B}
\author{Jan Robrade,
        J{\"u}rgen H.M.M. Schmitt}

\affiliation{Hamburger Sternwarte, University of Hamburg, Gojenbergsweg 112, 21029 Hamburg, Germany}

\shorttitle{X-rays from $\alpha$~Cen}
\shortauthors{J. Robrade \& J. Schmitt}

\abs{We report on the coronal activity cycles of our stellar neighbors $\alpha$\,Centauri A/B. The binary has been monitored with XMM-Newton since 2002 to study the long-term evolution of coronal activity evolution in X-rays. The solar analog $\alpha$\,Cen~A was clearly detected early in the program, but virtually faded away from XMM's detectors view around 2005. After remaining nearly a decade in a state of coronal weakness, we now detect a clear re-brightening of its corona. The secondary $\alpha$\,Cen B dominates the X-ray emission at most times and more than a full cycle is covered for this star. A new X-ray maximum was observed around 2012 that is again followed by gentle dimming over the recent years. The temporal evolution of the X-ray emission can be well understood, in analogy to the 11~year solar-cycle, by coronal activity cycles with different amplitudes and periods operating in both stars.}

\begin{document}

\maketitle

\section{Introduction}

The well know binary $\alpha$~Centauri A/B system consist
of a G2 (A) and K1 (B) star at a distance of 1.3~pc that orbit each other with a period of 86~yr. The M~dwarf Proxima Centauri is in common proper motion with $\alpha$~Cen A/B, but outside the FOV of the presented data and not discussed here. Together they make up the stellar system closest to the Sun. 
The $\alpha$~Cen~system is with an age of about 5~Gyr slightly older than the Sun, $\alpha$~Cen A/B are both slow rotators ($P_{\rm rot} = 30-40$~d) and show overall low activity levels of $\log L_{\rm X}/L_{\rm bol} \approx -6 \dots -7$. The secondary is slightly more active than the primary and largely dominates the X-rays emission of the system.

Here we give an update on results from our XMM-Newton X-ray monitoring campaign that has been ongoing since 2003 with typically one or two observations taken per year; the Chandra HRC-I joined the monitoring in 2005 after the major X-ray dimming of $\alpha$~Cen~A.
We continue our long-term study presented \cite{rob12} and extend the analysis with observations performed in the years 2012\,--\,2016.
For details regarding the used analysis methods and references we refer to several related publications; for the XMM campaign see \cite{rob05, rob12}, for an analysis of the Chandra data see \citet{ayr09, ayr14}, a long-term view in X-ray/FUV/UV of $\alpha$~Cen~B is presented in \cite{dew10}. 

\subsection{X-ray data}

For the purpose of this study XMM-Newton and Chandra complement each other quite well.
The XMM data is used to generate images, light curves and medium resolution spectra, results are derived from the EPIC MOS and pn detectors and refer to the 0.2\,--\,2.0~keV range.
The $\alpha$~Cen binary is quite well resolved in early XMM data, 
but the dwindling separation between both components due to their orbital motion, from about 14'' to 4'' over the campaign, complicates the analysis of the individual components in later data.
In contrast, the Chandra HRC-I has a superb spatial resolution of $\lesssim 1 ''$, but is virtually a photon counting machine (0.08\,--\,10.0~keV) and 
we use the measured count rates for the comparison with XMM. 
Three additional observations were taken with the HRC-S, where we use the zeroth order; the Chandra HRC-S/LETGS and XMM/RGS grating spectra are not discussed here.
Calibration differences at the lowest energies between XMM/EPIC and Chandra/HRC as well as the temperature dependence of the conversion factors introduce an uncertainty of about a factor of two in the regime of 1~MK coronal plasma, here relevant for the activity minimum phase of $\alpha$~Cen~A.
The conversion from HRC-S/LETGS zeroth order to HRC-I counts also depends on plasma temperature, the adopted range refers to average plasma temperatures of 1\,--\,5~MK;
a description of the instruments is given at {\it http://www.cosmos.esa.int/web/xmm-newton} and {\it http://cxc.harvard.edu}.

\begin{figure}
\centering
\includegraphics[width=0.85\linewidth]{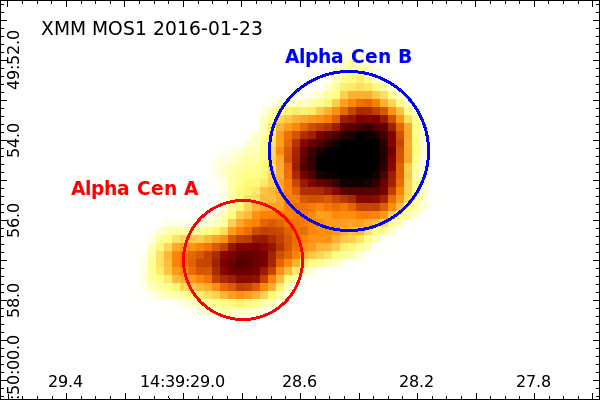}
\caption{The $\alpha$~Centauri A/B system in soft X-rays as seen with the MOS1 detector in January 2016.}
\label{im16}
\end{figure}

\section{The $\alpha$~Cen system in X-rays}

An X-ray image of the $\alpha$~Cen~A/B system taken in January 2016 with XMM MOS1 is shown in Fig.~\ref{im16}. The image is smoothed and uses a linear scaling, the circular regions mark the positions of the two binary components. Both are clearly detected, even if at the current separation of about 4'' the binary is not fully resolved by XMM. 

The reappearance $\alpha$~Cen~A in XMM images is mainly due a coronal re-brightening over the last years and to increasing emission from hotter plasma that produces sufficiently hard X-ray photons. This behavior is similar to the solar one, where active regions and larger spot groups are rare to absent during cycle minima. The previous minimum of $\alpha$~Cen~A had a duration of nearly a decade. While the $\alpha$~Cen~A minimum appears quite long and deep when compared to typical solar minima, prolonged inactive phases are also known for the Sun.
Further, over the last years $\alpha$~Cen~B is approaching a new coronal minimum and the reduced contrast between the components facilitates the detection of the primary.

\begin{figure}[t]
\centering
\includegraphics[width=0.95\linewidth]{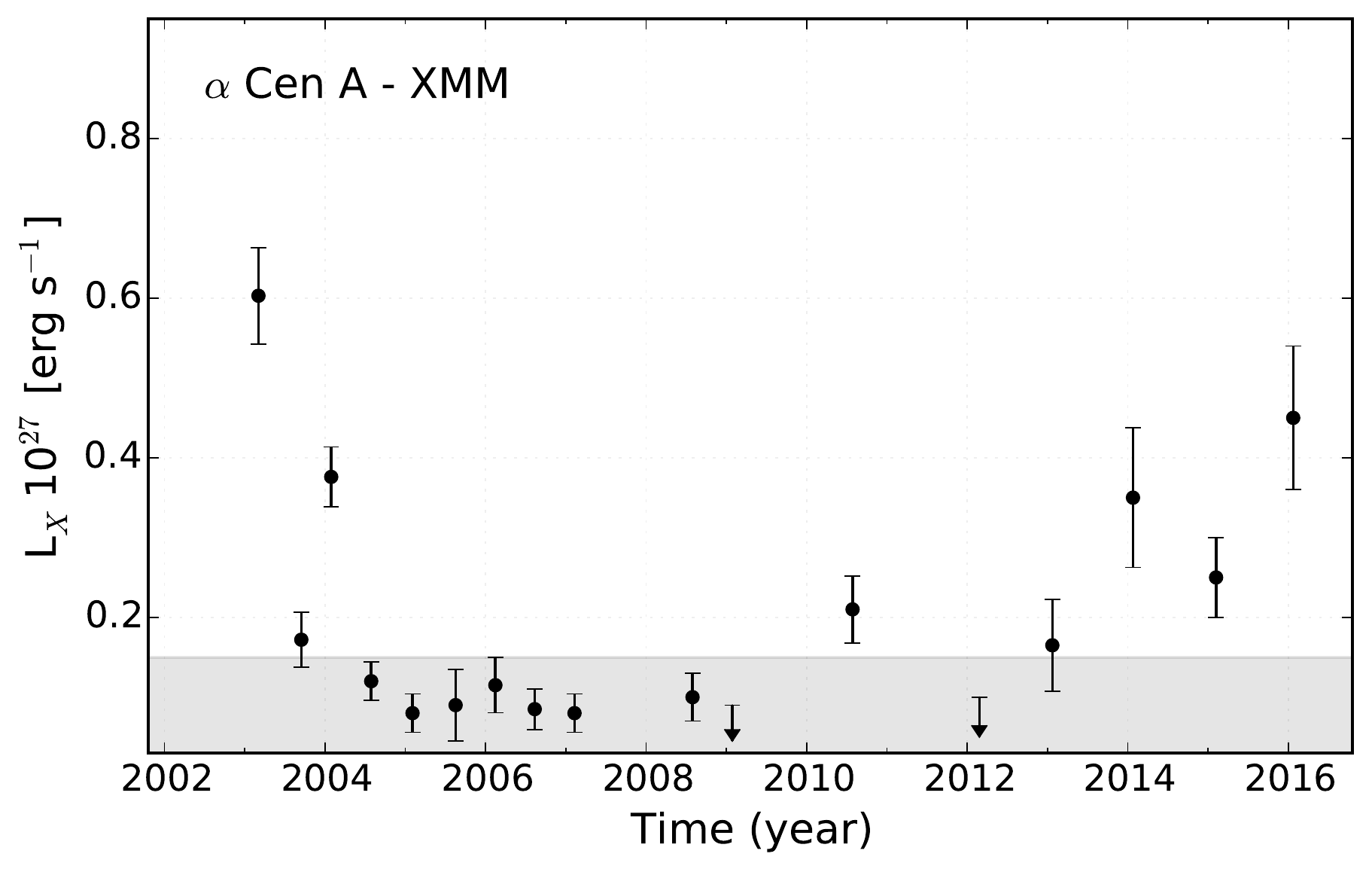}
\caption{Long-term X-ray luminosity of $\alpha$~Centauri A, XMM data at 0.2\,--\,2.0~keV. The grey shades region marks marginal detections, downward arrows denote upper limits.}
\label{lca}
\end{figure}

\begin{figure}[t]
\centering
\includegraphics[width=0.95\linewidth]{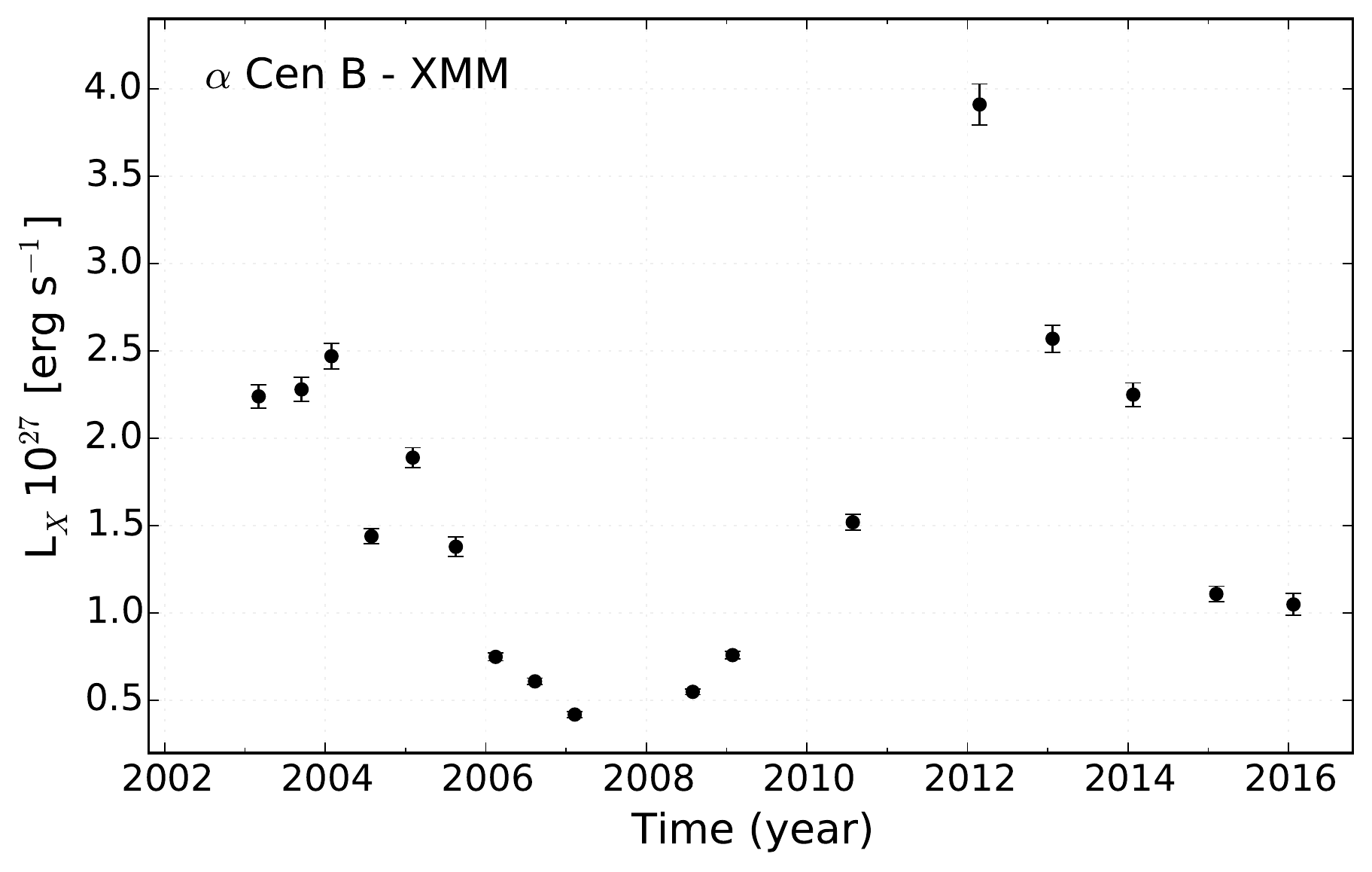}
\caption{Long-term X-ray luminosity of $\alpha$~Centauri B, XMM data at 0.2\,--\,2.0~keV. The evolution shows cyclic behavior.}
\label{lcb}
\end{figure}

The long-term evolution of the X-ray luminosity of $\alpha$~Cen~A (G2) and $\alpha$~Cen~B (K1) derived from the XMM data is shown in Fig.~\ref{lca} and Fig.~\ref{lcb} respectively, in both cases a more or less clear cyclic behavior is present or indicated.

\subsection{The activity cycle of $\alpha$~Cen~A}

The primary $\alpha$~Cen~A was clearly detected in the first XMM observation in 2003 with $L_{\rm X} \approx 6 \times 10^{26}$~erg\,s$^{-1}$, but subsequently showed a strong and quite rapid X-ray dimming. Early in the campaign between 2003 and 2005, the coronal emission measure, especially those of the hotter plasma in the temperature range of about 2\,--\,10~MK, decreased dramatically. As a consequence,
the star virtually faded from XMM's view and remained extremely X-ray faint from about 2005\,--\,2012, indicating a pronounced activity minimum. 
At activity minimum with its lower X-ray brightness and the corresponding cooler plasma temperatures even non-detections with XMM are present.
During this phase detections are often marginal and the derived X-ray luminosities are prone to larger uncertainties, also due to the presence of the secondary, here indicated by the grey shaded region in Fig.~\ref{lca}.
In the following years $\alpha$~Cen~A became continuously X-ray brighter and the XMM observations show a reappearance of hotter coronal emission that is associated with active regions, a typical characteristic of coronal activity cycles. 

\begin{figure}[t]
\centering
\includegraphics[width=0.95\linewidth]{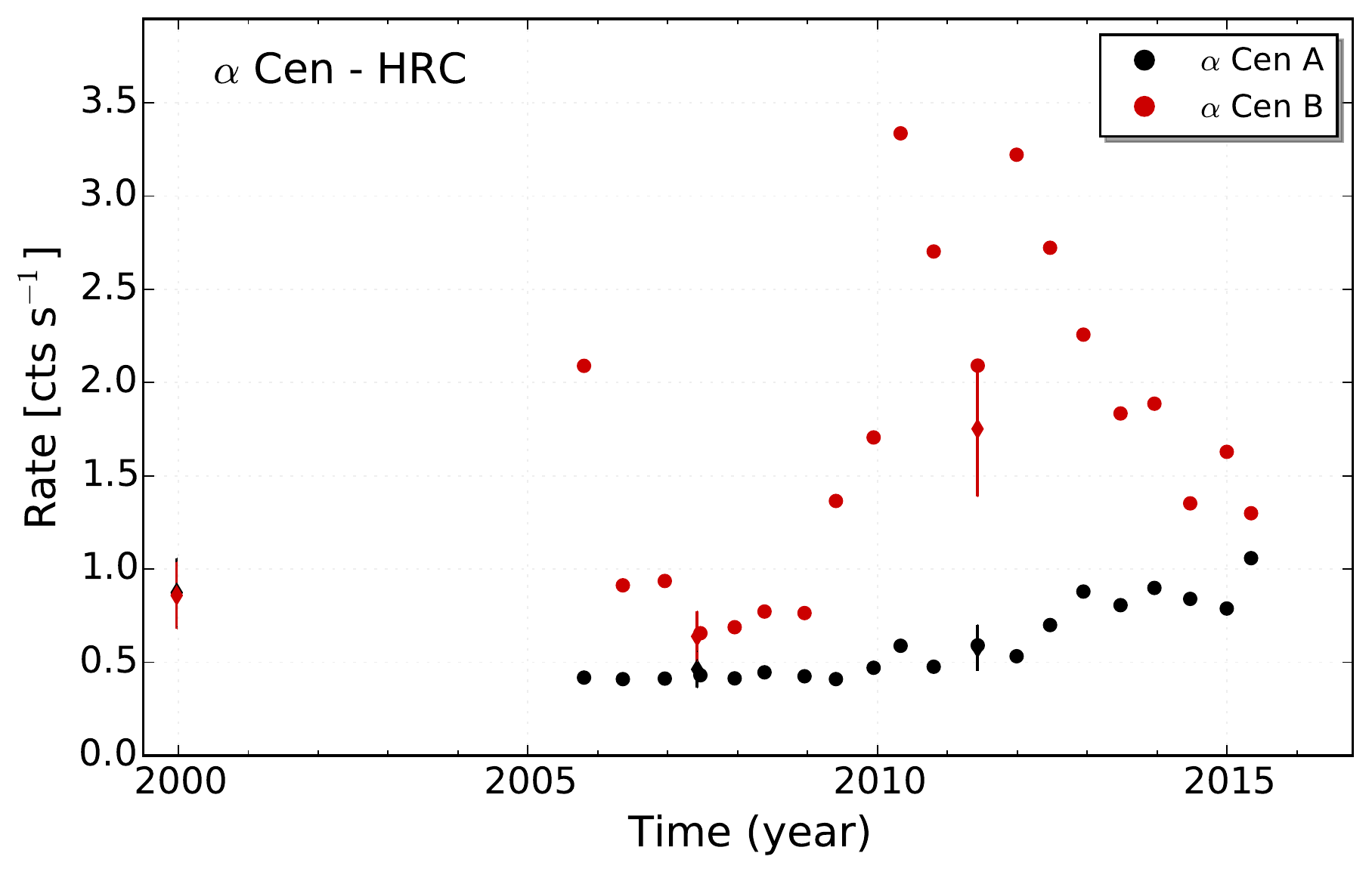}
\caption{Long-term X-ray light curve of $\alpha$~Cen A (black) and B (red) from Chandra. Count rates (1.5'' radius) from HRC-I (circles) and HRC-S/LETGS zeroth order scaled (diamonds).}
\label{lcc}
\end{figure}

When comparing the long-term lightcurve from XMM with those from Chandra, where $\alpha$~Cen~A remained visible in the HRC at all times (see Fig.~\ref{lcc}), overall consistent trends are seen. The HRC is sensitive to much softer X-ray emission and the better PSF of Chandra allows a clear separation between the components. Direct light curve comparison comparison and various emission line studies show that the very cool corona at 1~MK exhibits significantly less cyclic variability than the hotter coronal components, resulting in cycle amplitudes that are strongly energy dependent.

The combination of all X-ray data indicates that the last activity maximum occurred at the turn or early in the millennium and that the activity cycle likely has a period of about 15\,--\,20~years. In the obtained XMM data we see an X-ray luminosity ratio $L_{\rm X max}/L_{\rm Xmin}$ of nearly one order of magnitude, albeit larger uncertainties are present and a clear maximum has not been covered yet. Observations are ongoing and
one has to await future observations to really cover and quantify a full coronal cycle for $\alpha$~Cen~A.

\subsection{The activity cycle of $\alpha$~Cen~B}

In contrast, already one and a half coronal cycles are covered for $\alpha$~Cen~B, the corresponding period is $P_{\rm act} \approx 8-9$~yr. The XMM data include two consecutive maxima and one clear minimum as can be seen in Fig.~\ref{lcb}. The cycle amplitude described by the X-ray luminosity ratio $L_{\rm X max}/L_{\rm Xmin}$ is about 6\,--\,9, depending on the ex- or inclusion of the most extreme data point from 2012.
The activity maxima of 2003/2004 and 2011/2012 appear similar in peak flux, but individual exposures taken during the second maximum show further enhancements in X-ray brightness by 60\,--\,70~\%. By including the Chandra observations (see Fig.~\ref{lcc}) a better temporal sampling is achieved, that reveals significant brightness variations during the recent maximum. 
These variations are not present in the XMM data of the 2003/2004 maximum, however this maximum occurred before Chandra joined monitoring and may be purely by chance.  
From 2013 on $\alpha$~Cen~B's activity is declining once again, extrapolating from the existing data the next minimum is expected to occur around 2017. 

To constrain the evolution of coronal properties over the activity cycle, \cite{rob12} performed
 a spectral analysis the XMM data taken during the maximum and minimum state. The spectral models show similar temperature components, but an order of magnitude decline in emission measure for the hotter plasma at 3\,--\,10~MK compared to a more moderate decline by a factor of a few at cooler temperatures of about 1\,--\,2~MK. Interpreting the plasma components roughly as quiescent and active/flaring emission, this trend again reflects the behavior of the Sun over its activity cycle.

\subsubsection{$\alpha$~Cen~B at coronal maximum}

\begin{figure}[t]
\centering
\includegraphics[width=0.95\linewidth]{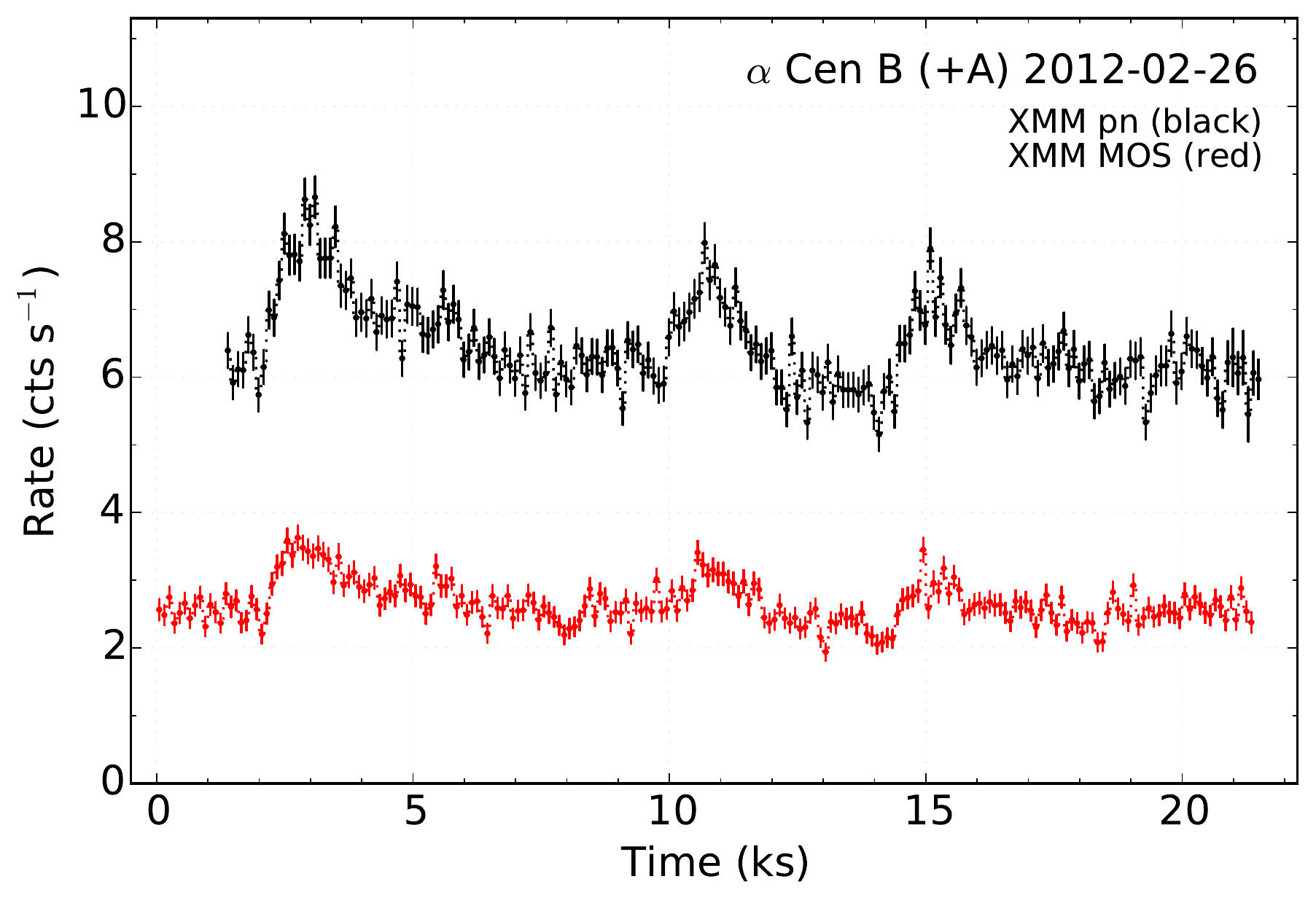}
\caption{XMM light curves of '$\alpha$~Cen~B' in early 2012 in the 0.2\,--\,2.0~keV band, here the secondary strongly dominates the X-ray emission of the system.}
\label{2012lc}
\end{figure}

The new XMM data cover the recent coronal maximum of $\alpha$~Cen~B and include an exceptional active phase around 2012. During the XMM observation the X-ray lightcurve exhibits frequent activity and a series of smaller flaring events as shown in Fig.~\ref{2012lc}.
The lightcurves are derived from the sum of both components, but are strongly dominated by $\alpha$~Cen~B that contributes $\gtrsim 90$~\% to the observed X-ray events as determined from the spatial photon distribution. Similarly, enhanced flux levels of $\alpha$~Cen~B are also repeatedly observed in Chandra exposures taken in the years 2010-2012, these often show comparable variability on timescales of hours.

From the spectral modelling of the 2012 XMM data we derive for $\alpha$~Cen~B an X-ray luminosity of $L_{\rm X} \approx 4 \times 10^{27}$~erg\,s$^{-1}$, the brightest quasi-quiescent state in more than 30 years of X-ray observations. The spectrum is slightly harder than during the 'normal' activity maximum state as for example observed in 2003/2004 and has a stronger contribution from 8~MK plasma that indicates the presence of pronounced active regions. However, overall the emission measure distribution is still dominated by 1\,--\,4~MK plasma and with an average temperature of about $T_{\rm X} = 2.5$~MK, reminiscent of a mildly active star.

\section*{Acknowledgments}
{JR acknowledges support from DLR under 50QR0803. Based on observations obtained with Chandra and XMM-Newton, used XMM data: 0045340901 (2003-03-04),
0045341001 (2003-09-15),
0045341101 (2004-01-29),
0045340401 (2004-07-29),
0143630501 (2005-02-01),
0143630201 (2005-08-17),
0202611201 (2006-02-15),
0202611301 (2006-08-13),
0202611401 (2007-02-09),
0550060901 (2008-07-30),
0550061001 (2009-01-27),
0654550801 (2010-07-29),
0670320301 (2012-02-26),
0690800301 (2013-01-22),
0720581101 (2014-01-24),
0741700301 (2015-02-06),
0760290301 (2016-01-23).}

\bibliographystyle{cs19proc}

\end{document}